\documentclass[12pt,a4paper]{article}
\usepackage[utf8x]{inputenc}
\usepackage{jheppub}

\renewcommand{\vec}[1]{\mathbf{#1}}
\renewcommand{\[}{\begin{equation}\begin{aligned}}
\renewcommand{\]}{\end{aligned}\end{equation}}

\renewcommand{\d}{\mathrm{d}}
\newcommand{\dd}{\hat{\mathrm{d}}}
\newcommand{\del}{\hat{\delta}}
\newcommand{\ket}[1]{| #1 \rangle}
\newcommand{\bra}[1]{\langle #1 |}

\renewcommand{\Re}{\operatorname{Re}}

\DeclareMathOperator{\sign}{sign}

\newcommand{\maxwell}{\phi}
\newcommand{\rootKerr}{$\sqrt{\text{Kerr}}$}

\title{A worldsheet for Kerr}
\author[1,2,3]{Alfredo Guevara,}
\author[4]{Ben Maybee,}
\author[5]{Alexander Ochirov,}
\author[4]{Donal O'Connell,}
\author[6]{and Justin Vines}

\affiliation[1]{Center for the Fundamental Laws of Nature, Harvard University, Cambridge, MA 02138, USA}
\affiliation[2]{Society of Fellows, Harvard University, Cambridge, MA 02138, USA}
\affiliation[3]{Black Hole Initiative, Harvard University, Cambridge, MA 02138, USA}
\affiliation[4]{Higgs Centre for Theoretical Physics, School of Physics and Astronomy, The University of Edinburgh, EH9 3FD, Scotland, UK}
\affiliation[5]{Mathematical Institute, University of Oxford,
Andrew Wiles Building, Radcliffe Observatory Quarter,
Woodstock Road, Oxford, OX2 6GG, UK}
\affiliation[6]{Max Planck Institute for Gravitational Physics (Albert Einstein Institute), Am M\"uhlenberg~1, Potsdam 14476, Germany}

\emailAdd{aguevaragonzalez@fas.harvard.edu}
\emailAdd{b.maybee@ed.ac.uk}
\emailAdd{ochirov@maths.ox.ac.uk}
\emailAdd{donal@ed.ac.uk}
\emailAdd{justin.vines@aei.mpg.de}

\abstract{
We show that the Newman-Janis shift property of the exact Kerr solution can be interpreted in terms of a worldsheet effective action. This
holds both in gravity, and for the single-copy \rootKerr~solution in electrodynamics. At the level of equations of motion, we show that the Newman-Janis
shift holds also for the leading interactions of the Kerr black hole. These leading interactions are conveniently described using chiral classical equations of motion
with the help of the spinor-helicity method familiar from scattering amplitudes.
}

\begin{document}
\maketitle
\addtocontents{toc}{\protect\setcounter{tocdepth}{1}}

\section{Introduction}
\label{sec:intro}

The Newman-Janis (NJ) shift~\cite{Newman:1965tw} is a remarkable exact property of the Kerr solution~\cite{Kerr:1963ud} describing a stationary spinning black hole in general relativity.
It relates the Kerr solution to the simpler non-spinning Schwarzschild solution.
One way to understand this property is to consider the Weyl curvature spinor $\Psi$. With appropriate coordinates, the NJ shift is
\[
\Psi^\text{Kerr}(x) = \Psi^\text{Schwarzschild}(x + i a) \,,
\]
where $a$ is the spin of the Kerr metric.
In other words, the Kerr solution looks like a complex translation
of the Schwarzschild solution \cite{Newman:2002mk}.

Precisely the same phenomenon occurs for the electromagnetic
\rootKerr~solution~\cite{Newman:1965tw},
which double-copies to the Kerr metric \cite{Monteiro:2014cda}.
In this case, it is the Maxwell spinor $\maxwell$ that undergoes a shift:
\[
\maxwell^{\sqrt{\text{Kerr}}}(x) = \maxwell^\text{Coulomb}(x + i a) \,.
\]
Therefore, \rootKerr~is a kind of complex translation of the Coulomb solution.

A partial understanding of this NJ phenomenon is available in the context of scattering amplitudes.
Although amplitudes are rooted in quantum field theory,
in recent years there has been an explosion of interest
in applications of amplitudes to classical physics
\cite{Neill:2013wsa,Bjerrum-Bohr:2013bxa,Vaidya:2014kza,Bjerrum-Bohr:2014zsa,Bjerrum-Bohr:2017dxw,Cachazo:2017jef,Guevara:2017csg,Cheung:2018wkq,Kosower:2018adc,Cristofoli:2019neg,Arkani-Hamed:2019ymq,Maybee:2019jus,Bjerrum-Bohr:2019kec,delaCruz:2020bbn,Mogull:2020sak,Cheung:2020gbf,Mougiakakos:2020laz,delaCruz:2020cpc}.
This is motivated by the applicability of the results for elastic
\cite{Bjerrum-Bohr:2018xdl,Guevara:2018wpp,Chung:2018kqs,Bern:2019nnu,KoemansCollado:2019ggb,Brandhuber:2019qpg,Emond:2019crr,Guevara:2019fsj,Bern:2019crd,Chung:2019duq,Damgaard:2019lfh,Burger:2019wkq,Aoude:2020onz,Bern:2020gjj,Chung:2020rrz,Cheung:2020gyp,Cristofoli:2020uzm,Bern:2020buy,Parra-Martinez:2020dzs,AccettulliHuber:2020oou,DiVecchia:2020ymx,Bjerrum-Bohr:2020syg}
and inelastic
\cite{Luna:2017dtq,Laddha:2018rle,Laddha:2018myi,Laddha:2018vbn,Sahoo:2018lxl,Bautista:2019tdr,Laddha:2019yaj,Saha:2019tub,Cristofoli:2020hnk,A:2020lub,Sahoo:2020ryf,Bonocore:2020xuj,Gonzo:2020xza} 
gravitational scattering of two massive particles
to gravitational-wave physics
\cite{Damour:2016gwp,Damour:2017zjx,Antonelli:2019ytb,Siemonsen:2019dsu,Kalin:2019rwq,Kalin:2019inp,Huber:2019ugz,Kalin:2020mvi,Cheung:2020sdj,Haddad:2020que,Kalin:2020fhe,Kalin:2020lmz,Bern:2020uwk,Aoude:2020ygw,Huber:2020xny},
as well as by the search for the underlying principles
behind the double-copy relationship between
gauge theory and gravity
\cite{Bern:2008qj,Bern:2010ue,Monteiro:2014cda,Johansson:2015oia,Luna:2016due,Goldberger:2016iau,Luna:2016hge,Goldberger:2017ogt,Li:2018qap,Shen:2018ebu,Plefka:2018dpa,Plefka:2019hmz,Johansson:2019dnu,Bautista:2019evw,Bern:2019prr,Plefka:2019wyg,Haddad:2020tvs,Carrasco:2020ywq,White:2020sfn}.
Recently, Arkani-Hamed, Huang and Huang~\cite{Arkani-Hamed:2017jhn}
pointed out a special class of three-point amplitudes
for a massive particle of spin $s$ emitting a gauge boson or graviton.
In a classical context, amplitudes for minimally coupled massive particles of large (classical) spin $s \gg 1$ are simple exponential factors
times the amplitudes for scalar
particles~\cite{Guevara:2018wpp,Guevara:2019fsj,Arkani-Hamed:2019ymq}.
In particular, the \rootKerr~three-point amplitude
describing classical electromagnetic interactions of a minimally coupled, massive, charged particle with a classical spin length $a$ is
\[
\mathcal{A}_{3,+}^{\sqrt{\text{Kerr}}} =
e^{-k \cdot a} \mathcal{A}_{3,+}^{\text{Coulomb}} \,.
\label{eq:rootKerrAmp}
\]
Here $\mathcal{A}_{3,+}^{\text{Coulomb}} = -2Q(p\cdot\varepsilon_k^+)$ is the usual QED amplitude for a scalar of charge $Q$ and momentum $p$ absorbing a photon with positive-helicity polarisation vector $\varepsilon_k^+$. Similarly, the gravitational three-point amplitude for a massive particle is
\[
\mathcal{M}_{3,+}^{{\text{Kerr}}} =
e^{-k \cdot a} \mathcal{M}_{3,+}^{\text{Schwarzschild}} \,,
\]
in terms of the ``Schwarzschild'' amplitude for a scalar particle
interacting with a positive-helicity graviton of momentum $k$.
A straightforward way to establish
the connection of these amplitudes to spinning black holes
\cite{Guevara:2018wpp,Chung:2018kqs,Guevara:2019fsj}
is to compute the impulse on a scalar probe at leading order in the Kerr background \cite{Arkani-Hamed:2019ymq}.
The calculation can be performed
using classical equations of motion on the one hand, and using scattering amplitudes and the KMOC
formalism~\cite{Kosower:2018adc,Maybee:2019jus,delaCruz:2020bbn}
on the other.
A direct comparison of the two approaches makes it evident
\cite{Guevara:2019fsj,Arkani-Hamed:2019ymq}
that the NJ shift of the background
is captured by the exponential factors $e^{\pm k\cdot a}$.

This connection between the NJ shift and scattering amplitudes
suggests that the NJ shift should extend beyond the exact Kerr solution
to the \emph{interactions} of spinning black holes.
Indeed, it is straightforward to scatter two Kerr particles
(by which we mean massive particles with classical spin lengths
$a_1$ and $a_2$)
off one another using amplitudes.
The purpose of this paper is to investigate
the classical interpretation of this fact.
To do so, we turn to the classical effective theory describing the worldline interactions of a Kerr particle \cite{Porto:2005ac,Porto:2006bt,Porto:2008tb,Steinhoff:2015ksa,Levi:2015msa}.
We will see that the NJ property endows this worldline action
with a remarkable two-dimensional worldsheet structure.
The Newman-Janis story emerges via Stokes'
theorem on this worldsheet with boundary and indeed persists
for at least the leading interactions.
We will see that novel equations of
motion, making use of the spinor-helicity formalism in a purely classical context, allow us to make the shift manifest in the leading interactions.

Our effective action is constructed only from the information in the three-point amplitudes. At higher orders, information from four-point
and higher amplitudes (or similar sources) is necessary to fully specify the effective action. Therefore our action is in principle supplemented
by an infinite tower of higher-order operators. We may hope, however, that the worldsheet structure may itself constrain the allowed higher-dimension operators. 

As applications of our methods, we use a generalisation of the Newman-Janis shift \cite{Talbot:1969bpa} to introduce magnetic charges (in electrodynamics) and NUT parameters (in gravity) for the particles described by our equations of motion.
As an example, we compute the leading impulse on a probe particle with mass,
spin and NUT charge moving in a Kerr-Taub-NUT background.
The charged generalisation of the NJ complex map can similarly be connected to the behaviour of three-point amplitudes in the classical limit \cite{Moynihan:2019bor,Huang:2019cja,Chung:2019yfs,Moynihan:2020gxj,Emond:2020lwi,Kim:2020cvf}, and we will reproduce results recently derived from this perspective~\cite{Emond:2020lwi},
furthermore calculating the leading angular impulse
(i.e. the change in spin during scattering) for the first time.

Our paper is organised as follows. We begin our discussion in the context of electrodynamics, constructing the effective action for a \rootKerr~
probe in an arbitrary electromagnetic background. In this case it is rather easy to understand how the worldsheet emerges. We discuss key
properties of the worldsheet, including the origin of the Newman-Janis shift, in this context. It turns out to be useful to perform the matching
in a spacetime with ``split'' signature $(+,+,-,-)$, largely because the three-point amplitude does not exist on-shell in Minkowski space.
The structure of the worldsheet is particularly simple
in split-signature spacetimes.
In section~\ref{sec:gr} we turn to the gravitational case, showing that the worldsheet naturally describes the dynamics of a spinning Kerr particle.
We discuss equations of motion in section~\ref{sec:spinorEOM},
focussing on the leading-order interactions which are not sensitive to
terms in the effective action which we have not constrained. In this section, we will see how useful the methods of spinor-helicity are for capturing the
chiral dynamics associated with the NJ shift, as well as magnetic charges. We finish our paper with a brief discussion.

\section{From amplitude to action}
\label{sec:rootKerrEFT}

We begin our story concentrating on the slightly simpler example of
the \rootKerr~particle in electromagnetism.
We wish to construct an effective action
for a massive, charged particle with spin angular momentum $S^{\mu\nu}$.
Building on the work of Porto, Rothstein, Levi and Steinhoff~\cite{Porto:2005ac,Porto:2006bt,Porto:2008tb,Levi:2014gsa,Levi:2015msa},
we write the worldline action as
\[
S = \int\!\d\tau \bigg\{ {-m}\sqrt{u^2} - \frac12 S_{\mu\nu} \Omega^{\mu\nu} - Q A \cdot u \bigg\} +  S_\text{EFT} \,,
\label{eq:fullSimpleAction}
\]
where $u^\mu$ and $\Omega^{\mu\nu}$ are the linear
and angular velocities,\footnote{We will be fixing $\tau$
to be the proper time,
so the velocity $u^\mu = \d r^\mu/\d\tau$ will satisfy $u^2=1  $.
The angular velocity can be defined through a body-fixed frame $e^a_\mu(\tau)$ on the worldline as
\[
\Omega^{\mu\nu}(\tau) = e^\mu_a(\tau) \frac{\d~}{\d\tau} e^{a\nu}(\tau) \,.
\label{eq:angMom}
\]
The tetrad allows us to pass from body-fixed frame indices $a, b, \ldots$
to Lorentz indices $\mu,\nu, \ldots $, as usual.
More details on spinning particles in effective theory
can be found in recent reviews~\cite{Porto:2016pyg,Levi:2018nxp}.}
and $S_\text{EFT}$ contains additional operators
coupling the spinning particle to the electromagnetic field.
We will be assuming the spin tensor to be transverse
according to the Tulczyjew covariant spin supplementary condition (SSC)
\[
S_{\mu\nu} p^\nu = 0 \,, \qquad \quad
p_\mu = -\frac{\partial L}{\partial u^\mu} = m u_\mu + {\cal O}(A) \,.
\label{eq:SSC}
\]
We can therefore relate the spin angular momentum to the spin pseudovector $a^\mu$ by
\[
a^{\mu} = \frac1{2p^2} \epsilon^{\mu\nu\rho\sigma} p_\nu S_{\rho\sigma} \qquad \Leftrightarrow \qquad
S_{\mu\nu} = \epsilon_{\mu\nu\rho\sigma} p^\rho a^\sigma \,.
\label{eq:spinEquivs}
\]
The effective action~\eqref{eq:fullSimpleAction} can be written independently of the choice of SSC, at the expense of introducing an additional term from minimal coupling \cite{Yee:1993ya,Porto:2008tb,Steinhoff:2015ksa}. This has played an important role in recent work pushing the gravitational effective action beyond linear-in-curvature terms \cite{Levi:2020kvb,Levi:2020uwu,Levi:2020lfn},
but for our present purposes a fixed SSC will suffice.
Note that any differences in the choice of the spin tensor $S_{\mu\nu}$
are projected out from the pseudovector $a^\mu$ by definition,
and it is the latter that will be central to our discussion.

We will only consider the effective operators in $S_\text{EFT}$
that involve one power of the electromagnetic field $A_\mu$,
which can be fixed by the three-point amplitudes.
Since these amplitudes are parity-even, 
the possible single-photon operators are
\[
S_\text{EFT} = Q\sum_{n=1}^\infty \int\!\d\tau \, u^\mu a^\nu
\big[ B_n  (a\cdot \partial)^{2n-2} {}^*\!F_{\mu\nu}(x)  + C_n (a \cdot \partial)^{2n-1} F_{\mu\nu}(x) \big]_{x=r(\tau)} \,.
\label{eq:introEFT}
\]
Notice that an odd number of spin pseudovectors is accompanied
by the dual field strength
\[
{}^*\!F_{\mu\nu} = \frac12 \epsilon_{\mu\nu\rho\sigma} F^{\rho \sigma} \,,
\]
while the plain field strength goes together with an even power of $a$.
By dimensional analysis,
the unknown constant coefficients $B_n$ and $C_n$ are dimensionless. 

\subsection{Worldsheet from source}

To determine the unknown coefficients,
we choose to match our effective action
to a quantity that can be derived directly from the three-point
\rootKerr~amplitude~\eqref{eq:rootKerrAmp}.
A convenient choice is the Maxwell spinor given by the amplitude for an incoming photon,
which is~\cite{Monteiro:2020plf}
\[
\phi(x) = -\frac{\sqrt{2}}{m} \Re \int\!\d\Phi(k) \, \del(k \cdot u) \, \ket{k} \bra{k} \, e^{- i k \cdot x } \mathcal{A}_{3,+} \,.
\]
In this expression, the integration is over on-shell massless phase space\footnote{Here and below we use the hat notation to absorb appropriate momentum-space factors of $2\pi$.}
\[
\d\Phi(k) = \dd^4k\, \del(k^2) \Theta(k^0) \equiv \frac{\d^4 k}{(2\pi)^4\!}\,(2\pi) \delta(k^2) \Theta(k^0) \,.\label{eq:phaseSpaceMes}
\]
This Maxwell spinor is defined in (2,2) signature.
Indeed, in Minkowski space, the only solution of the zero-energy condition $k \cdot u$ for
a massless, on-shell momentum
is $k^\mu=0$, so the three-point amplitude cannot exist on shell for non-trivial kinematics.
However, there is no such issue in (2,2) signature, which motivates analytically continuing from Minkowski space. (The spinor $\ket{k}$ is constructed from the on-shell null momentum $k$ as usual in spinor-helicity.)

In fact, the Newman-Janis shift makes it extremely natural for us to analytically continue to split signature even in the classical sense, without any consideration of three-point amplitudes.
The Maxwell spinor for a static \rootKerr~particle is explicitly
\[
\phi^{\sqrt{\text{Kerr}}}(x) = -\frac{Q}{4\pi} \frac{1}{(x^2 + y^2 + (z + i a)^2)^{3/2}} (x, y, z + ia) \cdot \boldsymbol{\sigma} \,.
\label{eq:phiKerrMink}
\]
In preparation for the analytic continuation $z= -iz'$,
we may choose to order the Pauli matrices
as $\boldsymbol{\sigma}=(\sigma_z, \sigma_x, \sigma_y)$.
Then the spinor structure in eq.~\eqref{eq:phiKerrMink} becomes real,
while the radial fall-off factor in the Maxwell spinor simplifies to
\[
\frac{1}{(x^2 + y^2 - (z - a)^2)^{3/2}} \,, \nonumber
\]
where we have dropped the prime sign of $z$.
In short, we have a real Maxwell spinor in (2,2) signature,
and the spin $a$ is now a real translation in the timelike $z$ direction.

We now analytically continue the action~\eqref{eq:introEFT}
by choosing the spin direction to become timelike.
In doing so, we also continue the component of the EM field in 
the spin direction, consistent with a covariant derivative
$\partial + i Q A$.
In split signature, it is convenient to rewrite the effective action
ansatz in terms of self- and anti-self-dual field strengths,
which we define as
\[
F^\pm_{\mu\nu}(x) = F_{\mu\nu}(x) \pm {}^*\!F_{\mu\nu}(x) \,.
\]
Our action then depends on a new set of unknown Wilson coefficients
$\tilde{B}_n$ and $\tilde{C}_n$:
\[
S_\text{EFT} = Q \sum_{n=0}^\infty \int\!\d\tau \, u^\mu a^\nu
\big[ \tilde{B}_n (a \cdot \partial)^n F_{\mu\nu}^+(x)
    + \tilde{C}_n (a \cdot \partial)^n F_{\mu\nu}^-(x)
\big]_{x=r(\tau)} \,.\label{eq:2,2actionUnmatched}
\]
To determine these coefficients we can match to the three-point amplitude by computing the Maxwell spinor for the radiation field sourced by the \rootKerr~particle, which we assume to have constant spin $a^\mu$ and constant proper velocity $u^\mu$. In (2,2) signature the exponential factor in eq.~\eqref{eq:rootKerrAmp} also picks up a factor $-i$, so we match our action to
\[
\phi(x) = -\frac{\sqrt{2}}{m} \Re\! \int\! \d\Phi(k) \, \del(k \cdot u) \, \ket{k} \bra{k} \, e^{- i k \cdot x } \mathcal{A}_{3,+}^\text{Coulomb} e^{i k\cdot a} \,.\label{eq:rKthreePointSpinor}
\]
The matching calculation is detailed in appendix~\ref{app:matching}, and determines the Wilson coefficients to be
\[
S_\text{EFT}  = Q \sum_{n=0}^\infty \int\!\d\tau\, u^\mu a^\nu
&\bigg[ {-\frac{(-a\cdot \partial)^n}{2(n+1)!}}  F_{\mu\nu}^+(x)
       + \frac{(a\cdot \partial)^n}{2(n+1)!} F_{\mu\nu}^-(x)
\bigg]_{x=r(\tau)} \\ 
= -\frac{Q}{2} \int\!\d\tau\, u^\mu a^\nu
&\bigg[ \left(\frac{e^{-a \cdot \partial} - 1}{-a \cdot \partial}\right) F_{\mu\nu}^+(x)
     - \left(\frac{e^{ a \cdot \partial} - 1}{a \cdot \partial}\right) F_{\mu\nu}^-(x)
\bigg]_{x=r(\tau)} \,.\label{eq:2,2actionMatchedCoefficients}
\]

So far, the Newman-Janis structure is hinted at by the translation
operators $e^{\pm a \cdot \partial}$ appearing in the effective action.
We can make this structure more manifest by writing the effective action equivalently as
\[
S_\text{EFT} & = -\frac{Q}{2}
\int\!\d\tau\!\int_0^1\!\d\lambda\, u^\mu a^\nu
\big[ e^{-\lambda 
(a \cdot \partial)} F_{\mu\nu}^+(x)
    - e^{\lambda (a \cdot \partial)} F_{\mu\nu}^-(x)
\big]_{x=r(\tau)} \\ &
= -\frac{Q}{2} \int\!\d\tau\!\int_0^1\!\d\lambda\, u^\mu a^\nu
\big[ F^+_{\mu\nu}(r - \lambda a) - F^-_{\mu\nu}(r + \lambda a) \big] \,.
\label{eq:seftStep}
\]
Our effective action is now an integral over a two-dimensional region
--- a worldsheet, rather than a worldline.

To see that this worldsheet is indeed connected to the Newman-Janis shift,
let us recover this shift for the Maxwell spinor.
First, we can read off the worldsheet current~$J^\mu$ from the action
$S_\text{EFT}-Q\!\int\!\d\tau A_\mu u^\mu = -\!\int\!\d^4x A_\mu J^\mu$.
Then the gauge field $A_\mu$ set up at a point~$x$ by this source
may be written as an integral of a Green's function $G(x-y)$ over the worldsheet:
\begin{align}
   A^\mu(x) & = \int\!\d^4y\,G(x-y) J^\mu(y) \\ &
    = Q\!\int\!\d\tau
      \bigg\{ u^\mu G(x-r)
           + \frac{1}{2}\!\int_0^1\!\d\lambda \Big(
             \big[ u^\mu (a \cdot \partial)
                 + 2\epsilon^{\mu\nu\rho\sigma}
                   u_\nu a_\rho \partial_\sigma
             \big] G(x-r+\lambda a) \nonumber \\ &
\qquad \qquad \qquad \qquad \qquad \qquad~\;\,\quad
           - \big[ u^\mu (a \cdot \partial)
                 - 2\epsilon^{\mu\nu\rho\sigma} u_\nu a_\rho \partial_\sigma
             \big] G(x-r-\lambda a) \Big)\!
      \bigg\} . \nonumber
\end{align}
The field strength follows by differentiation, after which contraction with $\sigma$ matrices yields the Maxwell spinor:
\[
\phi(x) = 2Q\!\int\!\d\tau\, \sigma_{\mu\nu}
u^{\nu} \partial^{\mu}\! \left[ G(x-r) - \int_0^1\!\d\lambda
\,(a \cdot \partial) G(x-r-\lambda a) \right] ,
\label{eq:NJbyintegration}
\]
where the first term comes from the non-spinning part of the action~\eqref{eq:fullSimpleAction}.
Now the $a\cdot\partial$ operator acting on the Green's function
can be understood as a derivative with respect to $\lambda$.
This produces a $\lambda$ integral of a total derivative,
which reduces to the boundary terms.
Cancelling the first term in eq.~\eqref{eq:NJbyintegration}
against the boundary contribution at $\lambda = 0$, we find simply that
\[
\phi(x) = 2Q\!\int\!\d\tau \, \sigma_{\mu\nu} u^\nu \partial^\mu  G(x-r-a)  \,.
\]
The Maxwell spinor depends only on the anti-self-dual part of the effective action, shifted by the spin length. The real translation in $(2,2)$ signature
is a result of this real worldsheet structure. We will shortly see that this structure persists for interactions.

\subsection{Worldsheet for interactions}

Let us analytically continue
the action~\eqref{eq:seftStep} back to Minkowski space:
\[
S_\text{EFT} & = \frac{Q}{2} \int\!\d\tau\!\int_0^1\!\d\lambda\,
u^\mu a^\nu \big[ i F^+_{\mu\nu}(r + i \lambda a)
                - i F^-_{\mu\nu}(r - i \lambda a) \big] \\ &
= Q \Re \int_\Sigma \! \d\tau \d \lambda \, iF^+_{\mu\nu}(r + i \lambda a) \, u^\mu a^\nu \,.
\]
Here the self-dual and anti-self-dual field strengths are
\[
F^\pm_{\mu\nu} = F_{\mu\nu} \pm i\,{}^*\!F_{\mu\nu}
= \pm \frac{i}{2} \epsilon_{\mu\nu\rho\sigma} F^{\pm\,\rho\sigma} \,,
\label{eq:rKwsaction}
\]
and $\Sigma=\{\tau\in(-\infty,\infty)\}\times\{\lambda\in[0,1]\}$
is the worldsheet.

\begin{figure}[t]
\center
\includegraphics[width = 0.5\textwidth]{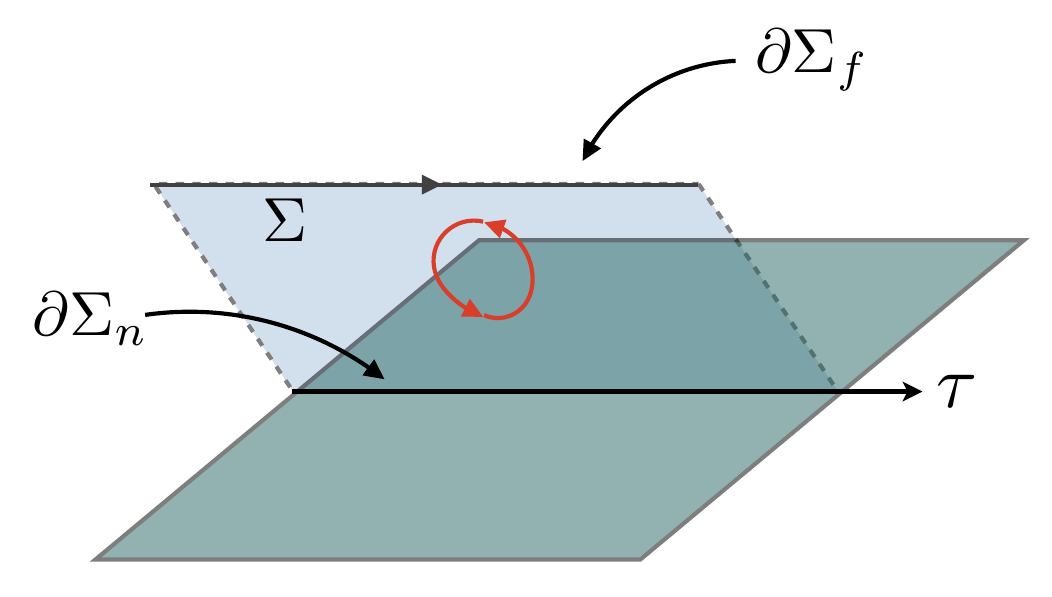}
\vspace{-3pt}
\caption{Geometry of the effective action: boundary $\partial \Sigma_\text{n}$ of the  complex worldsheet (translucent plane) is  fixed to the particle worldline in real space (solid plane).
\label{rootKerrIntegration}}
\end{figure}

Now that we are back in Minkowski space,
let us turn to the Newman-Janis structure of interactions.
Suppose that our spinning particle is moving
under the influence of an external electromagnetic field, generated by distant sources. The total interaction Lagrangian contains the worldsheet
term~\eqref{eq:rKwsaction} as well as the usual worldline minimal coupling:
\[
S_\text{int} = -Q \int_{\partial \Sigma_\text{n}}\!\!\!\d\tau A_\mu(r) u^\mu
+ Q \Re\! \int_\Sigma\!\d\tau \d \lambda \, iF^+_{\mu\nu}(r + i \lambda a) \, u^\mu a^\nu + \ldots \,,
\label{eq:sint}
\]
where $\partial \Sigma_\text{n}$ is the ``near'' boundary of the worldsheet, at $\lambda = 0$, as shown in figure~\ref{rootKerrIntegration}. We will similarly refer to the boundary at $\lambda = 1$ as the ``far'' boundary. The near boundary is the physical location of the object, while the far boundary is a timelike line embedded in the complexification of Minkowski space. We have also indicated the presence of unknown additional operators (involving at least two powers of the field strength) in the action by the ellipsis in eq.~\eqref{eq:sint}.

It is convenient to introduce a complex coordinate $z = r + i \lambda a$ on the worldsheet. In terms of this coordinate, we may write the two-form
\[
F^+(z) =
\frac{1}{2} F^+_{\mu\nu}(z) \d z^\mu \wedge \d z^\nu = i F^+_{\mu\nu}(z) (u^\mu + i \lambda \dot a^\mu) a^\nu \, \d \tau \wedge \d \lambda \,,
\]
where $\dot a^\mu = \d a^\mu / \d \tau$. Since in the absence of interactions the spin is constant, $\dot a$ must be of order $F$.
Therefore, we may rewrite our interaction action as
\[
S_\text{int} = -Q \int_{\partial \Sigma_\text{n}} \!\!\! A_\mu(r) \d r^\mu
+ \frac{Q}{2} \Re\! \int_\Sigma F^+_{\mu\nu}(z) \, \d z^\mu \wedge \d z^\nu + \ldots \,.
\label{eq:sintNicer}
\]
In doing so, we have redefined the higher-order operators indicated by the ellipsis.

When the electromagnetic fields appearing in the action~\eqref{eq:sintNicer} are generated by external sources,
both $F$ and ${}^*\!F$ are closed two-forms,
so we may introduce potentials $A$ and ${}^*\!A$
such that $F = \d A$ and ${}^*\!F = \d\,{}^*\!A$.
The dual gauge potential ${}^*\!A$ is related to $A$ by duality,
but this relationship need not concern us here:
we only require that both potentials exist in the vicinity
of the \rootKerr~particle.
Hence we may also write $F^+ = \d A + i\, \d\,{}^*\!A = \d A^+$.
Then the action~\eqref{eq:sintNicer} becomes
\[
S_\text{int} &
=-Q \int_{\partial \Sigma_\text{n}}\!\!A
+ Q \Re\!\int_\Sigma\!\d A^+ + \ldots \\ &
=-Q \int_{\partial \Sigma_\text{n}}\!\!A
+ Q \Re\!\int_{\partial \Sigma_\text{n}}\!\!A^+
- Q \Re\!\int_{\partial \Sigma_\text{f}}\!\!A^+ + \ldots\,,
\]
where the boundary consists of two disconnected lines (the far and near boundaries).
The orientation of the integration contour was set by $F^+$,
as depicted in figure~\ref{rootKerrIntegration}.

Now, notice that on the near boundary $z = r(\tau)$ is real.
Hence $\Re A^+ = A$,
so we are only left with the far-boundary contribution in the action:
\[
S_\text{int} = -Q \Re\!\int_{\partial \Sigma_\text{f}}\!\!A^+ + \ldots
=-Q \Re\!\int\!\d \tau \, u^\mu A_\mu^+(r + i a) + \ldots \,.
\label{eq:rtKerrIntShift}
\]
Thus we explicitly see that the interactions
of a \rootKerr~particle can be described with a Newman-Janis shift.
We will exploit this fact explicitly in section~\ref{sec:spinorEOM}.
Before we do, we turn to gravitational interactions.

\section{Spin and gravitational interactions}
\label{sec:gr}

As a step towards a worldsheet action for a probe Kerr in a non-trivial background, it is helpful to understand how to make the electromagnetic
effective action~\eqref{eq:rKwsaction} generally covariant. In a curved spacetime, we cannot simply add a vector $\lambda a$ to a point $r$.
To see what to do, let us reintroduce translation operators as in eq.~\eqref{eq:seftStep}. The worldsheet EFT term in Minkowski space is
\[
S_\text{EFT} &= Q \Re\! \int_\Sigma \! \d\tau \d\lambda\, 
i\, e^{i \lambda  \, a \cdot \partial} F^+_{\mu\nu}(x)\, u^\mu a^\nu \Big|_{x=r(\tau)}  \\
&= Q \Re \int_\Sigma \! \d\tau \d\lambda \, i
\sum_{n=0}^\infty \frac{1}{n!} (i \lambda \, a \cdot \partial)^n  F^+_{\mu\nu}(x) \, u^\mu a^\nu\Big|_{x=r(\tau)} \,.
\]
Now it is clear that a minimal way to make this term generally covariant is to replace the partial derivatives $\partial$ with covariant 
derivatives $\nabla$, so in curved space we have
\[
S_\text{EFT} &= Q \Re \!\int_\Sigma \! \d\tau \d\lambda \,
\sum_{n=0}^\infty \frac{1}{n!} (i \lambda \, a \cdot \nabla)^n F^+_{\mu\nu}(x) \, i u^\mu a^\nu  \Big|_{x=r(\tau)} \,.
\label{eq:rKeftCurvedStep}
\]

It is therefore natural for us to consider a covariant translation operator
\[
e^{i\lambda\,a \cdot \nabla} \equiv \sum_{n=0}^\infty \frac{1}{n!} (i \lambda \, a \cdot \nabla)^n \,.
\label{eq:explicitTranslation}
\]
This operator generates translations along geodesics in the direction $a$.
To see
why, note that the perturbative expansion of such a geodesic beginning at a point
$x_0$ in the direction $a$ with parameter $\ell$ is
\[
x^\mu(\ell) = x_0^\mu + \ell \, a^\mu - \frac{\ell^2}{2} \Gamma^\mu_{\nu\rho}(x_0) a^\nu a^\rho + \ldots \,.
\]
Now consider the perturbative expansion of a scalar function $f(x)$ along such a
geodesic. We have
\[
f(x(\ell)) &= f(x_0) + \ell \, a^\mu \partial_\mu f(x_0) + \frac{\ell^2}2 a^\mu a^\nu 
\left(
\partial_\mu \partial_\nu f(x_0) - \Gamma^\alpha_{\mu\nu}(x_0) \partial_\alpha f(x_0)
\right) 
+ \ldots \\
&=
f(x_0) + \ell (a \cdot \nabla) f(x_0) + \frac{\ell^2}2 (a \cdot \nabla) (a \cdot \nabla) f(x_0)
+ \ldots \\
&= e^{\ell \, a\cdot \nabla} f(x_0) \,.
\]

A traditional point of view on eq.~\eqref{eq:rKeftCurvedStep} is that the
operators only act on the two-form $F^+_{\mu\nu}$. However, we can alternatively
think of the operator acting on a scalar function $F^+_{\mu\nu} u^\mu a^\nu$, 
provided we extend the definitions of the velocity $u$ and the spin $a$
so that they become fields on the domain of the translation operator. We can simply
do this by parallel-transporting $u(r(\tau))$ 
and $a(r(\tau))$ along the geodesic beginning at $r(\tau)$ in the direction $a(\tau)$
(using the Levi-Civita connection). We denote these geodesics by $z(\tau, \lambda)$; 
explicitly,
\[
z^\mu(\tau, \lambda) = r^\mu(\tau) + i \lambda a^\mu(\tau) + \frac{\lambda^2}{2} \Gamma^\mu_{\nu\rho}(r(\tau)) a^\nu a^\rho + \ldots \,.
\]
(Notice that the translation operator~\eqref{eq:explicitTranslation} has parameter
$i \lambda$.)
The parallel-transported vectors, with initial conditions $a(z(\tau, 0)) = a(\tau)$ 
and $u(z(\tau, 0)) = u(\tau)$,
have the similar perturbative expansions
\[
u^\mu(z(\tau, \lambda)) &= u^\mu(\tau) - i \lambda \Gamma^\mu_{\nu\rho} (r(\tau)) 
a^\nu(\tau) u^\rho(\tau) + \ldots \,,\\
a^\mu(z(\tau, \lambda)) &= a^\mu(\tau) - i \lambda \Gamma^\mu_{\nu\rho} (r(\tau)) 
a^\nu(\tau) a^\rho(\tau) + \ldots \,.
\]
We now view the translation operator in eq.~\eqref{eq:rKeftCurvedStep}
as acting on the scalar quantity $F^+_{\mu\nu} u^\mu a^\nu$:
\[
e^{i \lambda a \cdot \nabla} F_{\mu\nu}(r(\tau)) a^\mu(\tau) u^\nu(\tau) 
= F_{\mu\nu}(z(\tau,\lambda)) a^\mu(z(\tau,\lambda)) u^\nu(z(\tau, \lambda)) \,.
\]
Expanding perturbatively to first order in $\lambda$, we have
\begin{align}
e^{i \lambda a \cdot \nabla} F_{\mu\nu}(r(\tau)) a^\mu(\tau) u^\nu(\tau)
&= \left(F_{\mu\nu}(r(\tau)) + i \lambda a^\rho \left(
\partial_\rho F_{\mu\nu} - \Gamma^\alpha_{\mu\rho} F_{\alpha\nu} 
-\Gamma^\alpha_{\nu\rho} F_{\mu \alpha} 
\right)\right) a^\mu u^\nu \nonumber \\
&= \left( F_{\mu\nu}(r(\tau)) + i \lambda a^\rho \nabla_\rho F_{\mu\nu}\right)a^\mu u^\nu \,.
\end{align}
The final expression is precisely the same as the picture in which 
the derivatives act only on the field strength: these are equivalent points of
view.

The worldsheet arises from interpreting the translation operators as genuine
translations. In curved space, the operators replace the straight-line sum 
$r+i a \lambda$ appearing in our action~\eqref{eq:rKwsaction} with the 
natural generalisation --- a geodesic in the direction $a$.\footnote{In general, 
these geodesics may become singular. We assume that such singularities do
not arise. If they were to arise, there would also be a divergence in the
interpretation of the EFT as an infinite sum of operators.}
We can express the curved-space effective action as
\[
S_\text{EFT} = Q \Re \!\int_\Sigma \! \d\tau \d \lambda \, iF^+_{\mu\nu}(z) \, u^\mu(z) a^\nu(z) \Big|_{z=z(\tau,\lambda)} \,.
\label{eq:rkCovActionVelocity}
\]
The surface $\Sigma$ is built up from the worldline of the particle, augmented by the geodesics in the direction $a$ for each $\tau$.

Note that, since we neglect higher-order interactions,
we may replace the velocity vector field $u(\tau,\lambda)$ 
in the action~\eqref{eq:rkCovActionVelocity}
with the similarly defined momentum field $p(\tau, \lambda)$.
Indeed, at $\lambda = 0$ the difference adds another order in the gauge field, as shown in eq.~\eqref{eq:SSC},
and this persists for $\lambda \neq 0$ after parallel translation along the geodesics.
Therefore, up to $F^2$ operators that we are neglecting,
the \rootKerr~action may be written as
\[
S_\text{EFT} = \frac{Q}{m} \Re\!\int_\Sigma\!\d\tau \d\lambda \, iF^+_{\mu\nu}(z) \, p^\mu(z) a^\nu(z) \Big|_{z=z(\tau,\lambda)} \,.
\label{eq:rkCovActionMomentum}
\]

We are now ready for the fully gravitational Kerr worldsheet action,
which is naturally motivated as a classical double copy of this covariantised worldsheet action.
Recalling that we should double-copy from non-Abelian gauge theory rather than electrodynamics, we promote the field strength to the Yang-Mills case:
\[
Q F^+_{\mu\nu}(z) ~\rightarrow~ c^A(z) F^{A+}_{\mu\nu}(z) \,,
\]
where $c^A(z(\tau,\lambda))$ is a vector in the colour space (generated by parallel transport from the classical colour vector of a particle,
as described by the Yang-Mills-Wong equations~\cite{Wong:1970fu}).
The double copy replaces colour by kinematics, so we anticipate a replacement of the form $c^A \rightarrow u^\mu$.
Moreover, to replace $F^{A}_{\mu\nu}$ we need an object with three indices,
antisymmetric in two of them,
for which the spin connection
\[
\omega_\mu{}^{ab} = e^b_\nu\, \nabla_\mu e^{a\nu} =
e^b_\sigma \big( \partial_\mu e^{a\sigma} + \Gamma^\sigma_{\mu\nu} e^{a\nu} \big)\label{eq:spinConnection}
\]
is the natural candidate.
Since it is defined via a derivative of the (body-fixed) spacetime tetrad~$e_a^\mu$, which is
a dimensionless quantity, on dimensional grounds the replacement should be
of the form $F^{A}_{\mu\nu} \to m \omega_\mu{}^{ab}$.
Indeed, we find that the correct worldsheet action for Kerr is
\[
S_\text{EFT} = \Re\! \int_\Sigma \d\tau \d \lambda \, i\,u_\mu(z) \, \omega^{+\mu}{}_{ab}(z) \, p^a(z) a^b(z) \Big|_{z=z(\tau,\lambda)} \,,
\label{eq:kerrws}
\]
where $\omega^+$ is a self-dual part of the spin connection,
defined explicitly by
\[
\omega^{+\mu}{}_{ab}(x) = \omega^{\mu}{}_{ab}(x) + i\,{}^*\!\omega^{\mu}{}_{ab}(x) \,, \qquad
{}^*\!\omega^{\mu}{}_{ab}(x) = \frac12 \epsilon_{abcd}\,\omega^{\mu\,cd}(x) \,.
\label{eq:omegaDefs}
\]
In writing these equations, we have extended the body-fixed frame $e_a = e_a^\mu \partial_\mu$ of vectors to every point of the complex worldsheet. We do so by parallel transport.
As usual, the frame indices $a, b, \cdots$ take values from 0 to 3,
and $\epsilon_{abcd}$ is the flat-space Levi-Civita tensor,
with $\epsilon_{0123} = + 1$.

\subsection{Flat-space limit}

We will shortly prove that the worldsheet term~\eqref{eq:kerrws} reproduces all single-curvature terms in the known effective action for a Kerr black hole in an arbitrary background~\cite{Porto:2006bt,Porto:2008tb,Levi:2015msa}. But first we wish to show that the term is non-trivial even in
flat space, and is in fact the standard kinetic term for a spinning particle in Minkowski space~\cite{Porto:2005ac} in that context.

In flat space and Cartesian coordinates, the worldsheet effective term~\eqref{eq:kerrws} is
\[
S_\text{EFT} = \Re \int\!\d\tau\!\int_0^1\!\d\lambda\,
i\,u_\mu(\tau)\,\omega^{+\mu}{}_{ab}(r + i \lambda a)\, p^a(\tau) a^b(\tau) \,, 
\label{eq:recoverSpinKineticStep}
\]
since the parallel transport of the vectors $u, p$ and $a$ is now trivial, and the geodesics reduce to straight lines. In flat space, 
the frame $e^a_\mu(\tau, \lambda)$ is also independent of $\lambda$, since it is
generated by parallel transport. 
Thus, the spin connection is $\lambda$-independent and the $\lambda$ integral in eq.~\eqref{eq:recoverSpinKineticStep} becomes trivial.

Given the $\lambda$ independence of the spin connection, we may write
\[
S_\text{EFT} & = \Re \int\!\d\tau\!\int_0^1\!\d\lambda\,i\,
u_\mu(\tau)\,\omega^{+\mu}{}_{ab}(r(\tau))\, p^a(\tau) a^b(\tau) \\ &
=-\!\int\!\d\tau\,
u_\mu(\tau)\,{}^*\!\omega^{\mu}{}_{ab}(r(\tau))\, p^a(\tau) a^b(\tau) \,.
\]
Recalling the definitions of the dual spin connection~${}^*\!\omega$ and the spin pseudovector~$a$, eqs.~\eqref{eq:omegaDefs} and \eqref{eq:spinEquivs}, we equivalently have
\[
S_\text{EFT} = - \frac12 \int\!\d\tau\,
u_\mu(\tau)\, \omega^{\mu}{}^{ab}(r(\tau))\, S_{ab}(\tau)
=- \frac12 \int\!\d\tau\, \Omega{}^{ab}(r(\tau))\, S_{ab}(\tau) \,.\label{eq:wsSpinKinematic}
\]
This is nothing but the spin kinetic term written in eq.~\eqref{eq:fullSimpleAction}.
In this way, we see that the worldsheet expression~\eqref{eq:kerrws} already describes the basic dynamics of spin.

\subsection{Single-Riemann effective operators}

It is now straightforward to recover the full tower of single-Riemann operators in the Kerr effective action. Returning to the full curved-space case, we may write our action~\eqref{eq:kerrws} as
\[
S_\text{EFT} & = \Re \!\int\!\d\tau\!\int_0^1\!\d\lambda\,
e^{i \lambda \, a \cdot \nabla} i\, u_\mu(\tau) \, \omega^{+\mu}{}_{ab}(r(\tau)) \, p^a(\tau) a^b(\tau) \\ &
= \Re \!\int\!\d\tau\, i\left(\frac{e^{i a \cdot \nabla}-1}{i a \cdot \nabla}\right) u_\mu(\tau) \, \omega^{+\mu}{}_{ab}(r(\tau)) \, p^a(\tau) a^b(\tau) \\ &
= \sum_{n=0}^\infty
\Re\!\int\!\d\tau\, \frac{(i a \cdot \nabla)^n}{(n+1)!} i\,u_\mu(\tau) \, \omega^{+\mu}{}_{ab}(r(\tau)) \, p^a(\tau) a^b(\tau) \,,
\]
where we performed the $\lambda$ integral and expanded the translation operator $e^{i a \cdot \nabla}$.
The leading contribution is again the spin kinetic term,
as a short computation demonstrates.
We also encounter an infinite series of higher-derivative contributions for $n \geq 1$. To express them in terms of the Riemann tensor,
we recall that it satisfies
\[
   R_{ab\,\mu\nu} = e^\alpha_a e^\beta_b R_{\alpha\beta\,\mu\nu}
    = -\nabla_\mu \omega_{\nu ab} + \nabla_\nu \omega_{\mu ab}
    + \omega_{\mu ac}\,\omega_{\nu}{}^c_{~\,b}
    - \omega_{\nu ac}\,\omega_{\mu}{}^c_{~\,b} \,.
\label{Connection2Riemann}
\]
Consistently omitting the quadratic in $\omega$ terms from the equation above, as well as the higher-order interaction contributions due to the difference between $p^a$ and $mu^a$,
we rewrite a typical effective operator as
\[
 & -\frac{1}{(n+1)!} \Re\!\int\!\d\tau\, u^\mu p^a a^b
   (i a \cdot \nabla)^{n-1} a^\nu \nabla_\nu \omega^+_{\mu ab}
   \big|_{x = r(\tau)} \\ & \quad
 =-\frac{1}{(n+1)!} \Re\!\int\!\d\tau\, p^a a^b u^\mu a^\nu
   (i a \cdot \nabla)^{n-1}
   \big[ R^+_{ab\,\mu\nu} + \nabla_\mu \omega^+_{\nu ab}
   \big]_{x = r(\tau)} + \ldots \\ & \quad
 =-\frac{m}{(n+1)!} \Re\!\int\!\d\tau\, u^a a^b a^\nu
   (i a \cdot \nabla)^{n-1}
   \bigg[ u^\mu R^+_{ab\,\mu\nu} + \frac{D~}{d\tau} \omega^+_{\nu ab}
   \bigg]_{x = r(\tau)} + \ldots \,.
\label{eq:EffOperatorSteps}
\]
Here $R^+_{ab\,\mu\nu}=R_{ab\,\mu\nu}+i\,{}^*\!R_{ab\,\mu\nu}$ is defined via the dualisation of the first two indices. Notice that in eq.~\eqref{eq:EffOperatorSteps} we treat 
the velocity $u$, momentum $p$ and spin $a$ as fields on the worldline, so that they
commute with the covariant derivative.

We may proceed by integrating the $D/\d \tau$ term by parts, after which it acts on factors of velocity and spin.
This generates curvature-squared (and higher) operators that we again neglect.
In this way, we arrive at the form of the leading interaction Lagrangian
\[
   S_\text{int} =-m\!\int\!\d\tau\,u^a a^b u^\mu a^\nu \Re \sum_{n=1}^\infty
   \frac{(i a \cdot \nabla)^{n-1}\!}{(n+1)!}
   \bigg[ R_{ab\,\mu\nu} + i\,{}^*\!R_{ab\,\mu\nu}
   \bigg]_{x = r(\tau)} + \ldots \,.
\]
Finally, separating the even and odd values of $n$ into two distinct sums
\[
S_\text{int} = m\!\int\!\d\tau
   \bigg[ & \sum_{n=1}^\infty \frac{(-1)^n}{(2n)!} (a \cdot \nabla)^{2n-2}
          R_{\alpha\beta\,\mu\nu} u^\alpha a^\beta u^\mu a^\nu \\ &
        -  \sum_{n=1}^\infty \frac{(-1)^n}{(2n+1)!}
          (a \cdot \nabla)^{2n-1}\,{}^*\!R_{\alpha\beta\,\mu\nu} u^\alpha a^\beta u^\mu a^\nu
   \bigg]_{x = r(\tau)} + \ldots \,,\label{eq:LSinteractions}
\]
one can verify that this reproduces the leading interactions of a Kerr black hole, as detailed in Appendix~\ref{app:kerrmatching} and discussed in detail by Levi and Steinhoff~\cite{Levi:2015msa}. 

It is interesting that the
worldsheet structure unifies the spin kinetic term with the
leading interactions of Kerr. The same phenomenon was observed
directly at the level of amplitudes in ref.~\cite{Chung:2018kqs}.

\section{Spinorial equations of motion}
\label{sec:spinorEOM}

Eq.~\eqref{eq:rtKerrIntShift} explicits displays a Newman-Janis shift for the leading interactions of the \rootKerr~solution. Now we take
a first look at the structure of the equations of motion encoding this shift. Since the Newman-Janis shift is chiral, we will find that it is very
convenient to describe the dynamics using the method of spinor-helicity, even in a fully classical setting. Our focus here will be to extract
expressions for observables from the equations of motion at leading order. Thus we are free to make field redefinitions, dropping total
derivatives which do not contribute to observables. We will also extend our work to magnetically charged objects, such as spinning dyons
and the gravitational Kerr-Taub-NUT analogue at the level of equations of motion.

We may write the leading order action for a \rootKerr~particle with trajectory $r(\tau)$ and spin $a(\tau)$ as
\[
S = - \int \d \tau \left( p \cdot \dot r(\tau) + \frac12 \epsilon(p, a, \Omega) + Q \Re u^\mu A_\mu^+(r + i a) \right) + \ldots \,.
\label{eq:rootKerrLeadingAction}
\]
Here and below we use the short-hand notation
\[
\epsilon(p, a, \Omega) = \epsilon_{\mu\nu\rho\sigma} p^\mu a^\nu \Omega^{\rho\sigma} \,,
\qquad \quad
\epsilon_\mu(a,b,c) = \epsilon_{\mu\nu\rho\sigma} a^\nu b^\rho c^\sigma \,.
\]
By varying with respect to the position $r(\tau)$ it is easy to determine that
\[
\frac{\d p^\mu}{\d \tau} &= Q \Re F^{+\mu\nu}(r+ia) u_\nu + \ldots
= \frac{Q}{m} \Re F^{+\mu\nu}(r+ia) p_\nu + \ldots \,.
\label{eq:rtKerrEOMmomentum}
\]
In the second equality, we replaced the velocity $u = \dot r$ with the momentum $p/m$, noting that the difference between the momentum and $mu$ is of 
order $F$. To obtain a similar differential equation for the spin $a^\mu$, it is helpful to begin by differentiating $a \cdot p = 0$, finding
\[
p_\mu \frac{\d a^\mu}{\d \tau} = p_\mu \Re \frac{Q}{m} F^{+\mu\nu}(r+ia) a_\nu \,.
\]
Based on this simple result, it is easy to guess that the spin satisfies
\[
\frac{\d a^\mu}{\d \tau} = \frac{Q}{m} \Re  F^{+\mu\nu}(r+ia) a_\nu + \ldots \,,
\label{eq:rtKerrEOMspin}
\]
and indeed a more lengthy calculation using the Lagrangian~\eqref{eq:rootKerrLeadingAction} confirms this guess.

Our expressions~\eqref{eq:rtKerrEOMmomentum} and~\eqref{eq:rtKerrEOMspin} for the momentum and spin have the same basic structure,
and are consistent with the requirements that $p^2$ and $a^2$ are constant while $a \cdot p = 0$. 
In the context of scattering amplitudes it has proven to be very convenient to introduce spinor variables describing similar momenta and spins.
Notice that there is nothing quantum about using spinor variables for momenta and spin: the momenta of particles in amplitudes need not be
small, and the spin can be arbitrarily large. We are simply taking advantage of the availability of spinorial representations of the Lorentz group.
A key motivation for introducing spinors in the present context is the chirality structure of eqs.~\eqref{eq:rtKerrEOMmomentum}
and~\eqref{eq:rtKerrEOMspin}, which hint at a more basic description using an intrinsically chiral formalism.

Following Arkani-Hamed, Huang and Huang~\cite{Arkani-Hamed:2017jhn} we introduce spinors $\ket{p_I}$ and $|p_I]$ with $I=1,2$
so that the momentum vector is
\[
p^\mu = \frac12 \epsilon^{IJ} \bra{p_J} \sigma^\mu | p_I] = \frac12 \bra{p^I} \sigma^\mu | p_I] \,.
\]
We raise and lower the little-group indices $I, J, \ldots$ with two-dimensional Levi-Civita tensors, as usual. The $\sigma^\mu$ matrices
are a basis of the Clifford algebra, and we use the common choice
\[
\sigma^\mu = (1, \sigma_x, \sigma_y, \sigma_z) \,.
\]
The little group of a massive momentum is $\text{SO}(3)$, so to construct the spin vector $a^\mu$ in terms of spinors we need only
form a little-group vector representation from little-group spinors. The vector representation of $\text{SO}(3)$ is the symmetric tensor product of two
spinors, so we will need to symmetrise little-group indices. Let $a^{IJ}$ be a constant symmetric two-by-two matrix; then
\[
a^\mu = \frac12 a^{IJ} \bra{p_J} \sigma^\mu | p_I] 
\]
is the spin vector. To understand how these expressions work, it may be helpful to work in a Lorentz frame $p^\mu = (\sqrt{p^2}, 0, 0, 0)$. Then
the spin is a purely spatial vector, so it is a linear combination of components in the $x$, $y$ and $z$ directions. Thus there is a basis of
three possible spins.
This is reflected in the three independent components
of the symmetric two-by-two matrix $a^{IJ}$. The algebra of the spinors
immediately guarantees that the spin $a$ and the momentum $p$ are orthogonal.

Given that we can always reconstruct the momentum and spin from the spinors, all we now need are dynamical equations for the spinors 
themselves. The leading-order spinorial equations of motion for \rootKerr~are
\[
\frac{\d}{\d \tau} \ket{p_I} &= \frac{Q}{2m} \maxwell(r(\tau) + i a(\tau)) \ket{p_I}\,, \\
\frac{\d}{\d \tau} |p_I] &=  \frac{Q}{2m} \tilde \maxwell(r(\tau) - i a(\tau)) |p_I] \,.
\label{eq:spinorEOMrtKerr}
\]
Notice that the evolution of the spinors is directly determined by the Maxwell spinor of whatever background the particle is moving in. The
NJ shift indicated explicitly in eqs.~\eqref{eq:spinorEOMrtKerr} is an explicit consequence of the shift~\eqref{eq:rtKerrIntShift} at the
level of the effective action. It is straightforward to recover the vectorial equations~\eqref{eq:rtKerrEOMmomentum} 
and~\eqref{eq:rtKerrEOMspin} from our spinorial equation using the familiar spinor-helicity methods.

To illustrate the use of spinorial methods, consider scattering two \rootKerr~particles off one another. We will compute both the leading impulse 
$\Delta p_1$ and the leading angular impulse $\Delta a_1$ on one of the two particles during the scattering event. These observables are
 easily obtained using the methods
of scattering amplitudes~\cite{Kosower:2018adc,Maybee:2019jus,Guevara:2019fsj,Arkani-Hamed:2019ymq}; here, spinorial equations of motion render the computation even simpler. We denote
the spinor variables for particle 1 by $\ket{1, \tau}$ and $|1, \tau]$, and 
similarly for particle 2; these spinors are explicitly functions of proper time. In a scattering event we denote the initial spinors
as $\ket{1} \equiv \ket{1, -\infty}$ (and similarly for $|1]$.) The
final outgoing spinors are then $\ket{1'} \equiv \ket{1, + \infty}$.

The impulses on particle 1 are given in terms of a leading order kick of the spinor $\ket{\Delta1} \equiv \ket{1, +\infty} - \ket{1, -\infty}$ as 
\[
\Delta p_1 &= 2 \epsilon^{IJ} \Re \ket{\Delta 1_{J}} [1_{I}| \,,\\
\Delta a_1 &= 2 a_1^{IJ} \Re \ket{\Delta 1_{J}} [1_{I}|  \,.
\label{eq:impulsesFromSpinors}
\]
(Notice that we are representing the impulses here as bispinors.) Thus we simply need to compute the kick suffered by the spinor of particle 1 to
determine \emph{both} impulses, in contrast to other methods available (including using amplitudes~\cite{Maybee:2019jus}.) By direct
integration of the spinorial equation~\eqref{eq:spinorEOMrtKerr} we see that this spinorial kick is
\[
\ket{\Delta 1_I} = \frac{Q_1}{2m_1} \int_{-\infty}^\infty\! \d \tau \,\phi(r_1 + i a_1) \, \ket {1_I} \,.
\label{eq:deltaSpinorStep}
\]
At this level of approximation, we may take the trajectory $r_1$ to be a straight line with constant velocity, and take the spin $a_1$ to 
be constant, under the integral. Notice that we evaluate the Maxwell spinor at the shifted position $r_1 + i a_1$ because of the Newman-Janis
shift property at the level of interactions.

To perform the integration we need the Maxwell spinor influencing the motion of particle 1. This is the field of the second of our two particles.
It is easy to obtain this field --- indeed, by the standard Newman-Janis shift of the field set up by particle 2, we need only shift the Coulomb
field of a point-like charge. The field is
\[
\phi(x) = 2iQ_2 \int\! \dd^4 k \, \del(k \cdot u_2) \frac{e^{-i k \cdot (x+i a_2)}}{k^2} \sigma_{\mu\nu} k^\mu u_2^\nu \,.
\label{eq:explicitMaxwell}
\]
Note the explicit NJ shift by the spin $a_2$: this is the shift of the background, in contrast to the shift through $a_1$ of 
eq.~\eqref{eq:deltaSpinorStep}. Of course, there is a pleasing symmetry between these shifts. Using the field~\eqref{eq:explicitMaxwell}
in our expression~\eqref{eq:deltaSpinorStep} for the change in the spinors of particle 1, we arrive at an integral expression for the spinor kick\,,
\[
\ket{\Delta 1_I} =  \frac{i Q_1 Q_2}{2m_1}\int \dd^4 k \, \del(k \cdot u_1)\del(k \cdot u_2) \frac{e^{-i k \cdot (b+i a_1+i a_2)}}{k^2} k^\mu u_2^\nu 
\sigma_{\mu\nu} \ket {1_I} \,,
\]
where $b$ is the impact parameter. 
This expression contains complete information about both the linear and angular impulses. For example, substituting into 
eq.~\eqref{eq:impulsesFromSpinors} we find that the angular impulse is
\[
\Delta a_1^\mu = \frac{Q_1 Q_2}{m_1} \Re\! \int\! \dd^4 k \, \del(k \cdot u_1)&\del(k \cdot u_2) \frac{e^{-i k \cdot (b+i a_1+i a_2)}}{k^2}
\\
&\times
\left(
ia_1 \cdot u_2 \, k^\mu - ik \cdot a_1 \, u_2^\mu + \epsilon^\mu(k, a_1, u_2) 
\right) \,.
\]

Spinorial equations of motion are also available for the leading order interactions of Kerr moving in a gravitational background. They are
\[
\frac{\d}{\d \tau} \ket{p_I} &= -\frac{1}{2} u^\mu \omega_\mu (r + i a) \ket{p_I}\,, \\
\frac{\d}{\d \tau} |p_I] &= - \frac{1}{2} u^\mu \tilde \omega_\mu (r - i a) |p_I] \,,
\label{eq:spinorEOMKerr}
\]
where the spin connection is written in terms of spinors:
\[
\omega_\mu \ket{p} = \omega_\mu{}^{ab} \sigma_{ab} \ket{p} \,.
\]
Using these spinorial equations of motion and a brief calculation in exact analogy with our \rootKerr~discussion above, it is straightforward
to recover the leading linear and angular impulse due to Kerr/Kerr scattering~\cite{Vines:2017hyw}. 

In fact we can go further and consider the generalisation of Kerr with NUT charge, corresponding in the stationary case to the Kerr-Taub-NUT
solution. It is known that NUT charge can be introduced by performing the gravitational analogue of electric/magnetic
duality \cite{Talbot:1969bpa}. Working at linearised level, this deforms the linearised spinorial equation of motion to
\[
\frac{\d}{\d \tau} \ket{p_I} &= -  \frac{e^{-i \theta}}{2} u^\mu \omega_\mu (r + i a) \ket{p_I}\,, \\
\frac{\d}{\d \tau} |p_I] &= -  \frac{e^{+i \theta}}{2} u^\mu \tilde \omega_\mu (r - i a) |p_I] \,,
\]
where $\theta$ is a magnetic angle. The particle described by these equations has mass $m \cos \theta$ and NUT parameter $m \sin \theta$.
Using these equations, and defining the rapidity $w$ by $\cosh w = u_1 \cdot u_2$, we find that the leading order impulse in a Kerr-Taub-NUT/Kerr-Taub-NUT scattering event is given by
\[
\Delta p_1^\mu = 4\pi G m_1 m_2 \Re\! \int\! \dd^4 k  \, \del(k \cdot u_1)&\del(k \cdot u_2) \frac{e^{-i k \cdot (b+i a_1+i a_2)}}{k^2} e^{i (\theta_2 -\theta_1)}
\\
\times&
\left(
i \cosh 2w \, k^\mu + 2 \cosh w \, \epsilon^\mu(k, u_1, u_2)
\right) \,,
\]
in agreement with a previous computation performed using scattering amplitudes~\cite{Emond:2020lwi}. It is also straightforward to compute
the angular impulse using these methods; we find that
\[
\Delta a_1^\mu = -4\pi G m_2 \Re\! \int\! \dd^4 k  \, \del(k \cdot u_1)\del(k \cdot u_2) \frac{e^{-i k \cdot (b+i a_1+i a_2)}}{k^2} e^{i (\theta_2 -\theta_1)} &
\\
\times
\Big(
i \cosh 2w \, \epsilon^\mu(k, u_1, a_1) - 2 \cosh w \, u_1^\mu \epsilon(k, u_1, u_2, a_1) & \\
- 2i a_1\cdot u_2 \cosh w\, k^\mu + ik\cdot a_1 \big(2 \cosh w\, u_2^\mu - u_1^\mu\big) &
\Big) \,.
\]

\section{Discussion}
\label{sec:discussion}

The Newman-Janis shift is often dismissed as a trick, without any underlying geometric justification. The central theme of our paper is that we
should rather view Newman and Janis's work as an important insight. The Kerr solution is simpler than it first seems, and correspondingly
the leading interactions of Kerr are simpler than they might otherwise be. It seems appropriate to place the NJ shift at the heart of our
formalism for describing the dynamics of Kerr black holes, thereby taking maximum advantage of this leading order simplicity.

Our spinorial approach to the classical dynamics of Kerr (and its electromagnetic single-copy, \rootKerr) makes it trivial to include the
spin (to all orders in $a$) in scattering processes. Computing the evolution of the spinors rather than the momenta and spin separately
reduces the workload in performing these computations, and is even more efficient in some examples than computing with the help of
scattering amplitudes. However, we only developed these equations at leading order. At higher orders, spinor equations of motion will certainly exist and be worthy of study.

We found that the effective action for Kerr has the surprising property that it can be formulated in terms of a two-dimensional worldsheet
integral instead of the usual one-dimensional worldline effective theory. This remarkable fact provides some kind of geometric basis for
the Newman-Janis shift, where it emerges using Stokes's theorem. Our worldsheet actions contain terms integrated over some boundaries,
and other terms integrated over the ``bulk'' two-dimensional worldsheet. This structure is also familiar from brane world scenarios, but is
obviously surprising in the context of Kerr black holes. In Minkowski space, this worldsheet is embedded in a complexification of spacetime,
in a manner somewhat reminiscent of other work on complexified worldlines; see ref.~\cite{Adamo:2009ru}, for example. However, 
our worldsheet seems to be a bit of a different beast: it is not a complex line, but rather a strip with two boundaries.

The worldsheet emerged in our work, built up from the physical boundary worldline and geodesics in the direction of spin. This construction
is very different from the sigma models familiar from string theory. The dynamical variables in our action are the ``near'' worldsheet coordinates,
the spin, and body-fixed frame. But perhaps these dynamical variables emerge from a geometric description more reminiscent of the picture for strings.

We do not know whether the worldsheet structure persists when higher-order operators, involving two or more powers of the Riemann
curvature (or electromagnetic field strength), are included. But we can certainly hope that the surprising simplicity of Kerr persists to higher
orders --- the computation of observables from loop-level amplitudes for particles with spin \cite{Guevara:2018wpp,Chung:2018kqs,Bern:2020buy,Aoude:2020ygw} certainly indicates that further progress can be made. Meanwhile precision calculations reliant on the effective action in \eqref{eq:fullSimpleAction} require including higher-dimension operators in the action  \cite{Levi:2020kvb,Levi:2020uwu,Levi:2020lfn}. It would be particularly interesting to investigate the symmetry structure of the Kerr worldsheet, with an eye towards placing symmetry constraints on the tower of possible higher-dimension operators.

Our work has touched on a family of solutions including Kerr and its magnetically charged analogue, Kerr-Taub-NUT, as well as a set
of related electromagnetic solutions. Taken together the parameters we have introduced number five: electric and magnetic charge, mass
and NUT parameter, and spin. These are most, but not all, of the parameters of the famous 
Plebanski-Demianski family~\cite{Debever:1971,Plebanski:1976gy}  of solutions,
which also includes a cosmological constant as well as an acceleration parameter. It would be interesting to see if our methods can
be generalised to include the remaining parameters.

We hope that our work is only the beginning of a programme to exploit the Newman-Janis structure of Kerr black holes to simplify their dynamics.

\section*{Acknowledgements}

We thank Tim Adamo, Lionel Mason, Ricardo Monteiro and Matteo Sergola.
AG has received support from NSF PHY-1707938 and from the Harvard Society of Fellows.
BM is supported by an STFC studentship ST/R504737/1.
AO's research is funded by the STFC grant ST/T000864/1.
DOC is supported by the STFC grant ST/P0000630/1.
AG and AO thank the Higgs Centre for Theoretical Physics for hospitality.

\appendix

\section{Split-signature matching calculation}
\label{app:matching}

Here we provide further details of the matching calculation which determined the \rootKerr~effective action Wilson coefficients in eq.~\eqref{eq:2,2actionUnmatched}. We will work exclusively in split signature, for which we adopt the conventions of \cite{Monteiro:2020plf}.

Our calculation hinges upon the field strength sourced by the particle, which is determined by the \rootKerr~worldline current $\tilde j^{\mu}(k)$. We assume that the particle has constant spin $a^\mu$ and constant proper velocity $u^\mu$. In solving the Maxwell equation we impose retarded boundary
conditions precisely as in~\cite{Monteiro:2020plf}, placing our observation point $x$ in the future with respect to one time coordinate $t^0$, but choosing the
proper velocity $u$ to point along the orthogonal time direction. It is useful to make use of the result
\[
\frac{1}{k^2_\text{ret}} = -i \sign k^0 \del(k^2) + \frac{1}{k^2_\text{adv}} \,,
\]
where the $\text{ret}$ and $\text{adv}$ subscripts indicate retarded and advanced Green's functions, respectively. Since the advanced Green's
function has support for $t^0 < 0$, we may simply replace
\[
\frac{1}{k^2_\text{ret}} = -i \sign(k^0) \del(k^2)  \,.\label{eq:2,2retardedProp}
\]
The field strength sourced by the current in split signature is therefore
\[
F^{\mu\nu}(x)  &= \int\!\dd^4k\, \sign(k^0) \del(k^2) \, k^{[\mu} \tilde j^{\nu]}\,e^{-ik\cdot x}\\
&= \int\!\dd^4k\, \left(\Theta(k^0) - \Theta(-k^0)\right) \del(k^2) \,  k^{[\mu} \tilde j^{\nu]}\,e^{-ik\cdot x}\,,
\]
where our convention for index antisymmetrisation includes no numerical factors. Notice that the appropriate integral measure is now precisely the invariant phase-space measure~\eqref{eq:phaseSpaceMes}; substituting the worldline current for our \rootKerr~effective action in eq.~\eqref{eq:2,2actionUnmatched}, evaluated on a leading-order trajectory, we thus have
\begin{multline}
F^{\mu\nu}(x) = 2Q \Re\! \int\!\d\Phi(k)\,\del(k\cdot u) \bigg\{k^{[\mu} u^{\nu]}\Big[1 + ia\cdot k \sum_{n=0}^\infty \left(\tilde B_n (ia\cdot k)^{n} + \tilde C_n (ia\cdot k)^{n}\right)\Big]\\
 + ik^{[\mu} \epsilon^{\nu]}(k,u,a) \sum_{n=0}^\infty\left(\tilde B_n(ia\cdot k)^{n} - \tilde C_n (ia\cdot k)^{n}\right)\bigg\} \, e^{-ik\cdot x}\,.\label{eq:unmatchedField}
\end{multline}

To match to the three-point \rootKerr~amplitude, we need to compute the Maxwell spinor $\phi$ and its conjugate, $\tilde \phi$. To do so, we introduce a basis of positive and negative helicity polarisation vectors $\varepsilon_k^{\pm}$. On the support of the delta function in~\eqref{eq:unmatchedField}, manipulations using the spinorial form of the polarisation vectors given in \cite{Monteiro:2020plf} then lead to
\[
k^{[\mu} u^{\nu]} \sigma_{\mu\nu} &= +\sqrt{2} \varepsilon_k^+\cdot u\, |k\rangle \langle k|\\
k^{[\mu} \epsilon^{\nu]}(k,u,a) \sigma_{\mu\nu} &= -\sqrt{2} a\cdot k\ \varepsilon_k^+\cdot u \,|k\rangle \langle k|\,.
\]
The latter equality relies upon the identity $k^{[\mu} \epsilon^{\nu\rho\sigma\lambda]} = 0$, and the fact that $\sigma_{\mu\nu}$ is self-dual in this signature. With these expressions in hand, it is easy to see that the Maxwell spinor has a common spinorial basis, and takes the simple form
\[
\phi(x) = 2\sqrt{2} Q \Re\! \int\! \d\Phi(k)\, \del(k\cdot u) e^{-ik\cdot x}\,& |k\rangle \langle  k| \,\varepsilon_k^+\cdot u  \\ &\times \left(1 + \sum_{n=0}^\infty 2 \tilde C_n (ia\cdot k)^{n+1}\right).\label{eq:unmatchedMaxwell}
\]
Recall from eq.~\eqref{eq:2,2actionUnmatched} that the Wilson coefficients $\tilde B_n$ and $\tilde C_n$ were identified with self- and anti-self-dual field strengths, respectively. Since a positive-helicity wave is associated with an anti-self-dual field strength, it is no surprise that the Maxwell spinor should depend only on this part of the \rootKerr~effective action. Fixing the $\tilde B_n$ coefficients requires the dual spinor, which is given by
\[
\tilde \phi(x) = -2\sqrt{2} Q \Re\! \int\! \d\Phi(k)\, \del(k\cdot u) e^{-ik\cdot x}\,& |k]\, [ k|\, \varepsilon_k^-\cdot u \\ &\times \left(1 + \sum_{n=0}^\infty 2 \tilde B_n (ia\cdot k)^{n+1}\right)\,.
\label{eq:unmatchedMaxwellDual}
\]
Here we have used that
\[
k^{[\mu} u^{\nu]} \tilde\sigma_{\mu\nu} &= -\sqrt{2} \varepsilon_k^-\cdot u\, |k]\,[ k|\\
k^{[\mu} \epsilon^{\nu]}(k,u,a) \tilde\sigma_{\mu\nu} &= -\sqrt{2} a\cdot k\ \varepsilon_k^-\cdot u \,|k]\,[ k|\,,
\]
recalling that $\tilde \sigma$ is anti-self-dual in split signature spacetimes.

It now only remains to match to the Maxwell spinors for the three-point~amplitude, as given in eq.~\eqref{eq:rKthreePointSpinor}. The scalar Coulomb amplitudes for photon absorption are just $\mathcal{A}_\pm = -2mQ\, u\cdot\varepsilon_k^\pm$, so for the \rootKerr~three-point amplitude
\[
\phi(x) &= + 2\sqrt{2}  Q \Re\! \int\! \d\Phi(k)\, \del(k\cdot u) e^{-ik\cdot x}\, |k\rangle \langle k| \, \varepsilon_k^+\cdot u \, e^{i k\cdot a}\,,\\
\tilde\phi(x) &= -2\sqrt{2}  Q \Re\! \int\! \d\Phi(k)\, \del(k\cdot u) e^{-ik\cdot x}\,  |k]\, [k|\, \varepsilon_k^-\cdot u\,e^{-i k\cdot a}\,.
\]
Expanding the exponentials and matching to eqs.~\eqref{eq:unmatchedMaxwell} and \eqref{eq:unmatchedMaxwellDual} then yields
\begin{equation}
\tilde B_n = \frac{(-1)^{n+1}}{2(n+1)!}\,, \qquad \quad
\tilde C_n = \frac{1}{2(n+1)!}\,,  
\end{equation}
which are the Wilson coefficients listed in eq.~\eqref{eq:2,2actionMatchedCoefficients}.

\section{Kerr matching calculation}
\label{app:kerrmatching}

Here we verify that the effective spin interactions
in eq.~\eqref{eq:LSinteractions} match to the leading interactions
for a Kerr black hole.
We start with the stress-energy tensor
\cite{Vines:2017hyw}
\[
\!\!\!\!T^{\mu\nu}_\text{Kerr}(x) = m\!\int\!\d\tau\;\!
      u^{(\mu} \exp(a*\partial)^{\nu)}_{~~\rho} u^\rho
      \delta^{(4)}(x-r(\tau)) \,, \qquad
   (a*b)^{\mu\nu} = \epsilon^{\mu\nu\alpha\beta} a_\alpha b_\beta \,,
\label{KerrEnergyTensor}
\]
which in the static case serves as an effective skeleton source
for the linearised Kerr solution
$g^\text{Kerr}_{\mu\nu} = \eta_{\mu\nu} + \kappa h^\text{Kerr}_{\mu\nu}$.
The trace-reversed form of the solution is explicitly
\[
   \bar{h}^\text{Kerr}_{\mu\nu}(x)
    = -u_{(\mu} \exp(a*\partial)_{\nu)}^{~~\rho} u^\rho\,
      \frac{\kappa m}{8\pi|\vec{x}|} \,, \qquad
   u^\mu = (1,\vec{0}) \,, \quad
   \kappa = \sqrt{32\pi G} \,.
\label{KerrLinSolution}
\]
Therefore, the corresponding interaction Lagrangian is
\begin{align}
   S^\text{Kerr}_\text{int} &
    = -\frac{\kappa}{2}\!\int\!\d^4x\,
      h^{\mu\nu}(x) T_{\mu\nu}^\text{Kerr}(x) + {\cal O}(\kappa^2) \\ &
    = -\frac{\kappa m}{2}\!\int\!\d\tau\,u_\mu \sum_{n=0}^\infty
      \bigg[ \frac{(a*\partial)^{2n}}{(2n)!}
           - \frac{(a*\partial)^{2n+1}}{(2n+1)!} \bigg]_{\!\nu}^{~\rho}
      u_\rho h^{\mu\nu} \bigg|_{x=r(\tau)}\!
    + {\cal O}(\kappa^2) \,. \nonumber
\end{align}
We can gradually trade powers of $(a*\partial)$
for powers of $(a\cdot\partial)$ by using
\[
   \big[(a*\partial)^2\big]_\nu^{~\,\rho} u_\rho
    = -u_\nu \big[(a\cdot\partial)^2 - a^2 \partial^2\big]
    + \big[a_\nu (a\cdot\partial) - a^2 \partial_\nu\big] (u\cdot\partial)
    + {\cal O}(\kappa) \,,
\]
where we have already discarded $u \cdot a = {\cal O}(\kappa)$.
Plugging this into the worldline action
and assuming harmonic gauge for simplicity,
we can also eliminate $\partial^2 h^{\mu\nu} = {\cal O}(\kappa)$.
Moreover, integration by parts of the proper-time derivatives
$(u\cdot\partial) h^{\mu\nu} = \d h^{\mu\nu}/\d\tau$
produces momentum or spin time derivatives,
which bring about additional powers of $\kappa h_{\mu\nu}$,
so we may neglect them as well.
By a repeated use of these steps, we obtain
\begin{align}
\label{KerrLinLagrangian}
   S^\text{Kerr}_\text{int}
    = -\frac{\kappa m}{2}\!\int\!\d\tau \sum_{n=0}^\infty
      \bigg[ & \frac{(-1)^n\!}{(2n)!} (a\cdot\partial)^{2n}
             u_\mu u_\nu h^{\mu\nu} \\ &
           - \frac{(-1)^n}{(2n+1)!} (a\cdot\partial)^{2n}
             u_\mu \epsilon_\nu(u, a, \partial) h^{\mu\nu}
      \bigg]_{x=r(\tau)}\!+ {\cal O}(\kappa^2) \,. \nonumber
\end{align}
Note that here the first two terms,
$-\kappa m/2 \big[ h_{\mu\nu} u^\mu u^\nu - u_\mu \epsilon_\nu(u, a, \partial) h^{\mu\nu} \big]$,
 should be interpreted \cite{Chung:2018kqs} as the linearised-gravity contributions
to the standard kinetic terms $-m\sqrt{u^2} - S_{ab} \Omega^{ab}/2$.
This can be seen by rewriting
\[
\Omega^{ab}(\tau) = u^\mu \omega_\mu{}^{ab} \big|_{x=r(\tau)}
= \frac{\kappa}{2} u^\mu \big[\partial^a h_\mu{}^b - \partial^b h_\mu{}^a\big]_{x=r(\tau)} + {\cal O}(\kappa^2) \,,
\]
where we have chosen the frame as $e_\mu^a = \delta_\mu^a + \kappa h_\mu{}^a/2$.

Meanwhile, from ref.~\cite{Levi:2015msa} it is well known that the leading effective spin interactions of any compact object can be written as
\[
S_\text{LS} = m\!\int\!\d\tau
   \bigg[ & \sum_{n=1}^\infty \frac{(-1)^n}{(2n)!} C_{\text{ES}^{2n}}
          (a \cdot \nabla)^{2n-2} R_{\alpha\beta\,\mu\nu}
          u^\alpha a^\beta u^\mu a^\nu \\ &
        + \sum_{n=1}^\infty \frac{(-1)^n}{(2n+1)!} C_{\text{BS}^{2n}}
          (a \cdot \nabla)^{2n-1} {}^*\!R_{\alpha\beta\,\mu\nu}
          u^\alpha a^\beta u^\mu a^\nu
   \bigg]_{x = r(\tau)} + \ldots\,,
\label{eq:EffectiveInteractions}
\]
where the dimensionless Wilson coefficients
$C_{\text{ES}^{2n}}$ and $C_{\text{BS}^{2n}}$ specify
the gravitational multipole moments of a given classical body.
To determine these coefficients for Kerr black holes,
we need to reduce the action~\eqref{eq:EffectiveInteractions}
to the linearised form of eq.~\eqref{KerrLinLagrangian}.
Recalling that at leading order any proper time derivatives
$(u\cdot\partial) h^{\mu\nu}$ can be neglected,
we find that in the worldline action the Riemann tensor may be replaced by
\[
R_{\alpha\beta\,\mu\nu} u^\alpha a^\beta u^\mu a^\nu &
~\Rightarrow~ -\frac{\kappa}{2} (a\cdot\partial)^2 h_{\mu\nu} u^\mu u^\nu
+ {\cal O}(\kappa^2)\,,\\
{}^*\!R_{\alpha\beta\,\mu\nu} u^\alpha a^\beta u^\mu a^\nu &
~\Rightarrow~ -\frac{\kappa}{2} u_\mu \epsilon_\nu(u, a, \partial) (a\cdot\partial) h^{\mu\nu}
+ {\cal O}(\kappa^2)\,.
\]
Therefore, the linearised leading spin interactions are
\begin{align}
\label{eq:EffectiveLinInteractions}
S_\text{LS} = -\frac{\kappa m}{2}\!\int\!\d\tau
   \bigg[ & \sum_{n=1}^\infty \frac{(-1)^n\!}{(2n)!} C_{\text{ES}^{2n}}
          (a \cdot \partial)^{2n} h_{\mu\nu} u^\mu u^\nu \\ &
        + \sum_{n=1}^\infty \frac{(-1)^n}{(2n+1)!} C_{\text{BS}^{2n}}
          (a \cdot \partial)^{2n} u_\mu \epsilon_\nu(u, a, \partial) h^{\mu\nu}
   \bigg]_{x = r(\tau)}\!
 + {\cal O}(\kappa^2) \,. \nonumber
\end{align}
Matching this to the linearised action~\eqref{KerrLinLagrangian},
we see that the Wilson coefficients for a Kerr black hole are
\[
C_{{\rm ES}^{2n}} = 1\,, \qquad \quad C_{{\rm BS}^{2n}} = -1\,.
\]
Substitution into \eqref{eq:EffectiveInteractions} then yields
the Kerr interactions listed in eq.~\eqref{eq:LSinteractions}.

\bibliographystyle{JHEP}
\bibliography{references}
\end{document}